\documentclass[manuscript]{acmart}
\usepackage{subcaption}
\usepackage{todonotes}
\usepackage{cleveref}
\usepackage{pgfplots}
\pgfplotsset{compat=1.18}

\AtBeginDocument{%
  }

\newcommand{\system}{\emph{HeyFriend Helper}}

\begin{document}

\title{HeyFriend Helper: A Conversational AI Web-App for Resource Access Among Low-Income Chicago Residents}

\author{Maddie Juarez}
\email{mjuarez4@luc.edu}
\orcid{0009-0002-6701-4469}
\authornotemark[1]
\affiliation{%
  \institution{Loyola University Chicago}
  \city{Chicago}
  \state{Illinois}
  \country{USA}
}

\author{Abha Rai}
\affiliation{%
  \institution{Loyola University Chicago}
  \city{Chicago}
  \country{USA}}
\email{arai4@luc.edu}

\author{Kristen E. Ravi}
\affiliation{%
  \institution{The University of Tennessee}
  \city{Knoxville}
  \country{USA}}
\email{kravi2@utk.edu}

\author{Margaret C. Delaney}
\affiliation{%
  \institution{Loyola University Chicago}
  \city{Chicago}
  \country{USA}}
\email{mdelan@luc.edu}

\author{Danny Olweean}
\affiliation{%
  \institution{Trellus}
  \city{Chicago}
  \country{USA}}
\email{dolweean@mytrellus.org}

\author{Eric Klingensmith}
\affiliation{%
  \institution{Independent Researcher}
  \city{Madison}
  \country{USA}}
\email{klingene@gmail.com}

\author{Swarnali Banerjee}
\affiliation{%
  \institution{Loyola University Chicago}
  \city{Chicago}
  \country{USA}}
\email{sbanerjee@luc.edu}

\author{Neil Klingensmith}
\affiliation{%
  \institution{Loyola University Chicago}
  \city{Chicago}
  \country{USA}}
\email{nklingensmith@luc.edu}

\author{George K. Thiruvathukal}
\affiliation{%
  \institution{Loyola University Chicago}
  \city{Chicago}
  \country{USA}
}

\renewcommand{\shortauthors}{Juarez et al.}

\begin{abstract}
  Low-income individuals can face multiple challenges in their ability to seek employment. Barriers to employment often include limited access to digital literacy resources, training, interview preparation and resume feedback. Prior work has largely focused on targeted social service or healthcare applications that address needs individually, with little emphasis on conversational AI-driven systems that integrate multiple localized digital resources to provide comprehensive support. This work presents HeyFriend Helper, a web-based platform designed to support low-income residents in Chicago through an interactive conversational assistant that provides personalized support and guidance. HeyFriend Helper integrates multiple tools, including resume building and feedback, interview practice, mindfulness and well-being resources, employment trend and career outcome information, language learning support, and location-based access to community services. This work represents an interdisciplinary collaboration between social work, computer science, and engineering that addresses the multifaceted needs of low-income individuals. The findings demonstrate the importance of career-readiness tools and conversational user interface (CUIs) in providing holistic support.
\end{abstract}




\maketitle

\section{Introduction}
Wealth disparity is the unequal distribution of assets and resources, shaped by an interaction of economic, social, and institutional factors that affect residents of large metropolitan areas. The social conditions for individuals with low incomes in Chicago are strongly shaped by interconnected social and neighborhood-level factors, including concentrated poverty, racial and economic segregation, unemployment, and limited access to essential resources such as healthy food options and transportation \cite{wadsworth_working_2012}. Factors influencing barriers to employment include limited access to jobs offering living wages, job training, education, reliable transportation, and structural and systemic inequities \cite{doi:10.1177/0020872815600509}. One way to address these barriers is through technological interventions, including web and mobile applications, that centralize informational resources, connect users to employment and certification opportunities, and provide accessible tools for skill development, mental well-being, and navigating local support services.

In this paper, we introduce a tool to address these barriers: \textbf{HeyFriend Helper}, a web application that provides personalized support through Conversational AI, employment resources, and certification help for low-income Chicago residents. This work is an interdisciplinary collaboration between social work, computer science, data science, nursing,engineering, and  our community partner, Trellus, that informed the system's design. The main contributions are the following:
\begin{itemize}
    \item We introduce a web-based platform that provides language learning capabilities, including basic phrase instruction and custom translation into English, to reduce language and accessibility barriers, which can be a challenge with this target population.
    \item The design of a holistic support system that combines mindfulness resources, location-based service discovery, and curated common questions to provide support that addresses the emotional and day-to-day needs of users.
    \item The development of career-readiness tools, including a resume builder with feedback and examples, as well as interactive interview practice to support employment preparation.
    \item The integration of an information-retrieval tool wrapped within a ChatGPT Assistant that provides real-time, personalized guidance and support, enabling adaptive user interaction.
    \item The first, to our knowledge, to combine elements of career-readiness, mindfulness, and holistic support targeted to low-income residents. 
    
\end{itemize}

\section{Background and Previous Work}
Prior work in this area has largely focused on developing tools that address specific needs within social service domains, rather than providing integrated, holistic support. For example, conversational user interfaces (CUIs) have been widely explored in healthcare settings to assist with screening, delivering targeted education, providing counseling, and support behavior-change interventions. 

Health-focused CUIs are generally used to guide patients, promote healthy behaviors, and offer digital interventions through conversational support, particularly in areas such as patient health education \cite{choi_user_2025}.
One such effort is a multi-chatbot mobile application designed to support sexual health education which uses rule-based and generative models to provide structured information \cite{sti}. These systems demonstrate how conversational agents can improve access to social services within healthcare, but are typically centered around a singular focus and have a limited use case. 

Furthermore, one study finds that conversational AI among low-income adults can meaningfully improve their quality of life \cite{kim-low}. The study demonstrated that CUIs can reduce loneliness and emotional distress by enabling conversation, music engagement (e.g., verbally requesting and listening to songs), and safety features (i.e., 911 calling capabilities), which contributed to improved mood, perceived security, and overall well-being. However, an important distinction is that Kim et al. focused on a smart speaker-based voice assistant designed to address low digital literacy, whereas our work incorporates both verbal and written conversational modalities, enabling users to engage across multiple communication preferences and literacy levels.

Similarly, prior work has explored how conversational user interfaces (CUIs) can support newcomers in navigating Canadian government services by evaluating how these systems provide information related to processes such as filing taxes, accessing employment and income supports, and obtaining immigration and settlement resources. These findings emphasize the role of CUIs as informational intermediaries that may reduce barriers related to digital literacy, language, and familiarity with government processes for immigrant and refugee populations \cite{taxes}.

While these tools can support users in navigating institutional systems and improving access to care and engagement, they tend to focus on single procedural goals or conditions rather than multi-domain social support needs. With the rapid rise of conversational large language models (LLMs), new opportunities have emerged for CUIs to support mental health and well-being, as well as time-sensitive issues, including employment readiness. Traditional avenues for personalized mental health support, such as therapy, can be financially costly, time-consuming, and difficult to access due to geographical constraints, creating additional obstacles for individuals seeking care \cite{brouwers_social_2020}. However, widely accessible LLMs enable conversational interactions that can be tailored to individual contexts, needs, and preferences, providing immediate forms of support. For example, Song et al. found that individuals reported feeling comforted through their interactions with LLM-based chatbots, using them as sources of advice, reassurance and support throughout routine conversations \cite{song2025typingcureexperienceslarge}. Previous work has also emphasized the importance of carefully designing conversational personas to enhance the user experience, including the use of avatars with voices and demographic-specific personals to enhance user experience and create more relatable interactions \cite{Zheng_2025}. Furthermore, there is considerable evidence to suggest that LLMs struggle to respond appropriately across diverse cultural contexts and often reflect biases embedded in their training data. Nevertheless, Aleem et al. demonstrate that ChatGPT can still function as a limited multicultural therapeutic agent and highlight where current models fall short in cultural understanding and contextual ability rather than functioning as a reliable standalone mental health resource \cite{aleem}. However, nuance and caution are necessary. 

Furthermore, Antico et al. \cite{antico2024unimibassistantdesigningstudentfriendly} introduced a Retrieval-Augmented Generation (RAG) chatbot that extends large language model capabilities by incorporating university-specific documents (e.g., maps, procedures) and links through OpenAI’s custom-GPT framework, demonstrating improved usability for students. The system enables the model to ground its responses in retrieved institutional materials rather than relying solely on pretrained knowledge, allowing students to access accurate, context-specific information within a single conversational interface. Without this retrieval layer, LLMs would struggle to answer contextually specific questions, such as those related to locations, policies, or procedures, that depend on institution-specific information not contained in general training data.

Together, these findings highlight an urgent need for culturally attuned tools that combine multiple forms of support, such as well-being resources, access to social services, and employment preparedness, into a single cohesive system.

\section{Design and Interface}

A primary design goal of our system was to keep features internal and directly accessible within the website in order to reduce the complexity associated with navigating to external tools or services. Prior work \cite{cognitive} suggests that increasing the number of interface elements and information sources can elevate cognitive load, thereby impacting usability. Motivated by this, we prioritized interface designs implemented locally in a centralized hub that support streamlined interaction and reduce unnecessary navigation overhead from navigating external resources.

Furthermore, a key contribution of this system is the development of a retrieval-augmented LLM designed to support users who may or may not be proficient in English and who may have low levels of digital literacy, literacy, and formal education. Addressing the needs of this population (i.e., low income individuals) presents several challenges, including that the model output remains interpretable, supportive, and culturally relevant while avoiding overly complex responses. 
To this end, our key design goals were as follows:
\begin{itemize}
    \item Reduce task complexity by embedding features directly within the site through API integration
    \item Organize functionality into modular tabs aligned with feature categories
    \item Use accordion-style dropdown components to present content progressively and maintain a simple interface
    \item Support multilingual accessibility through consistent 1-to-1 translation across all platform content
\end{itemize}
Each section of the informational website provides 1-to-1 translation of content in English, Spanish, Arabic, and French. These languages were chosen for discrete translation because they are common languages spoken in Chicago, IL and per recommendations from our community agency partner. Each section also includes a custom \textit{HeyFriend Helper}, a conversational AI tool embedded within each page to provide convenient support to users.

This website was developed as part of a short-term  study conducted in collaboration with a community partner. This study was designed and deployed to 25 intervention participants across a 4 week period in Chicago, IL, provided at a confidential location. The system was not publicly accessible and was deployed only for the duration of the study, remaining live for a limited period of several weeks. Access was restricted to study participants and research personnel. The system is deployed as a web application and can be accessed publicly \cite{heyfriendhelper}. No personally identifiable information was collected through the website. Any personal data associated with the study was collected through input surveys and was not stored in the site’s backend infrastructure. To further protect participant anonymity, no IP addresses or device identifiers were logged. The only data collected consisted of aggregated, non-identifiable usage metrics (e.g., buttons clicked, questions submitted, and tabs opened), which were used solely to assess overall system usage and interaction patterns.

\renewcommand{\arraystretch}{1.25}

\begin{table}[h!]
\centering
\begin{tabular}{|p{4cm}|p{9cm}|}
\hline
\textbf{Question} & \textbf{Predefined Response} \\ \hline

I feel guilty for not being able to care for or support all of my family members. What
should I do?
&
It’s completely normal to feel this way—many people do, especially when resources are limited.
Remember to be kind to yourself. You are doing the best you can. Please don’t blame yourself.
Even small efforts matter, and your presence, care, or communication can have a big impact. If
you're working or studying, remind yourself that you're building a foundation to help your loved
ones in the long run. Try to stay connected when possible—through regular phone calls,
messages, or video chats. Finding a local support group, church, or community center can also
give you strength and connection during this time. Also, if your loved one needs material support
you can contact 2-1-1 or share the number with your loved one. 2-1-1 is a 24/7 service that
connects people with resources and information across various support areas, including health and
social services. You can reach 2-1-1 by calling 2-1-1, texting your zip code to 898-211, or visiting
211MetroChicago.org. \\ \hline

What is a job interview like?
&
A job interview is a formal conversation between a job applicant and a potential employer,
designed to assess whether the applicant is a good fit for a specific role. It typically begins with
introductions and some small talk, followed by questions from the interviewer about the
candidate’s background, skills, experiences, and how they handle certain situations. Sometimes
job interviews will start with the question “Tell me about yourself”. This is a very broad question
so it is important to plan how you will answer it in a simple, straightforward way.
Job interviews will typically ask why you are interested in the job. Be prepared to have an
answer to this question. Depending on the position, the interview may include technical or
problem-solving tasks. Interviews can be conducted in person, over the phone, or via video calls.
Candidates are also usually given the opportunity to ask questions about the role or company.
The tone of the interview can vary from formal to more casual, but professionalism is always
important. After the interview, candidates often follow up with a thank-you note and wait to hear
about the next steps. Here is a template to incorporate your CV/Resume:
https://www.resume-now.com/build-resume/choose-template \\ \hline

\end{tabular}
\caption{Examples of Question-Response Pairs Used in RAG System}
\label{tab:shakila}
\end{table}

\begin{table}[h!]
\centering
\begin{tabular}{|p{4cm}|p{9cm}|}
\hline
\textbf{Iteration} & \textbf{Instruction Description} \\ \hline
1.0 & HeyFriend is an informational website for you to access resources on stress management, mindfulness, employment, resume writing, interviewing, practicing English, and a custom conversational chatbot. You are a helpful assistant who answers user questions clearly and concisely. \\ \hline
2.0 & You are a warm, supportive digital well-being and resource assistant serving low-income individuals and families in Chicago. Your goal is to provide clear, practical, multilingual guidance on mental wellness, mindfulness, employment support, resume/CV help, parenting tips, local resources, and social services. When appropriate, ask follow-up questions to better understand the user’s needs. Provide appropriate responses that are not too lengthy. Your mission is to make resources easier to access and to help reduce stress, improve wellness, and support financial stability for individuals and families in Chicago. \\ \hline
3.0 & You are a warm, supportive digital well-being and resource assistant serving low-income individuals and families in Chicago. Your goal is to provide clear, practical, multilingual guidance on mental wellness, mindfulness, employment support, resume/CV help, parenting tips, local resources, and social services. You respond kindly, non-judgmentally, and with empathy — always empowering users and encouraging self-care. You are familiar with community resources in Chicago (e.g., Trellus, 211, Chicago Workforce Centers, local food banks, schools, and healthcare options), and you are able to converse in multiple languages such as English, Spanish, French, and Arabic. When appropriate, ask follow-up questions to better understand the user’s needs. Provide answers in simple, accessible language, and offer translations in other languages when requested. Using the file provided with direct answers, suggest local tools, links, or hot-lines when additional support is needed. If a user asks a question mentioned in the document, answer with the exact response provided. Your mission is to make resources easier to access and to help reduce stress, improve wellness, and support financial stability for individuals and families in Chicago. \\ \hline
\end{tabular}
\caption{Progression of Instruction-tuning for HeyFriend Helper Conversational AI}
\label{tab:simple3x2}
\end{table}

\subsection{HeyFriend Helper Assistant}

We implement the conversational helper component of \textit{HeyFriend Helper} using OpenAI's ChatGPT Assistants framework \cite{OpenAI2024AssistantsAPI} designed for use with low-income residents in Chicago. \textit{HeyFriend Helper }is built on the \textit{gpt-4o-11-20-2024} model, the most recent archived model available for assistant deployment at the time of development. We use a RAG-like design in which a curated set of question-answer pairs serves as a controlled ground truth source for queries instead of full embedding-based retrieval pipeline. To ground the assistant in domain-specific knowledge and a ground-truth reference, we provided a curated file consisting of 20 question-answer pairs, seen in Table \ref{tab:shakila}. The questions and responses were developed using prior background research, LLM responses, and feedback/adjustment from the study team and the community agency partner to provide responses to the most commonly asked questions by the target population. These questions address common concerns related to navigation, family responsibilities, employment, personal well-being, and Chicago-specific information on navigating transportation systems and social services. 
The assistant was explicitly guided to first assess whether a user's query was included in the curated question set before generating a response. If an existing question was identified, \textit{HeyFriend Helper} was instructed to retrieve and present the corresponding response directly from the provided file. 

Furthermore, retrieval is enforced with verbatim responses. For all questions not explicitly stated in the question set, we leveraged the assistant's underlying language model's general conversational and contextual knowledge to generate responses. Table \ref{tab:simple3x2} summarizes the progressive instruction iterations used to guide the assistant's behavior, which became increasingly specific and targeted through iterative internal testing and refinement. Preliminary internal evaluations with the chatbot indicated that, without detailed description guides such as "\textit{provide concise responses}" or "\textit{ask follow-up questions when appropriate}", the assistant would often deviate from the subject matter and generate lengthy, generalized responses about social services or platform features. Moreover, to encourage responses to be RAG-like using our predefined questions as a ground truth reference, we enforced this behavior through the instruction "\textit{if a
user asks a question mentioned in the document, answer with the exact response provided.}"

\begin{figure}
    \centering
    \includegraphics[width=0.5\linewidth]{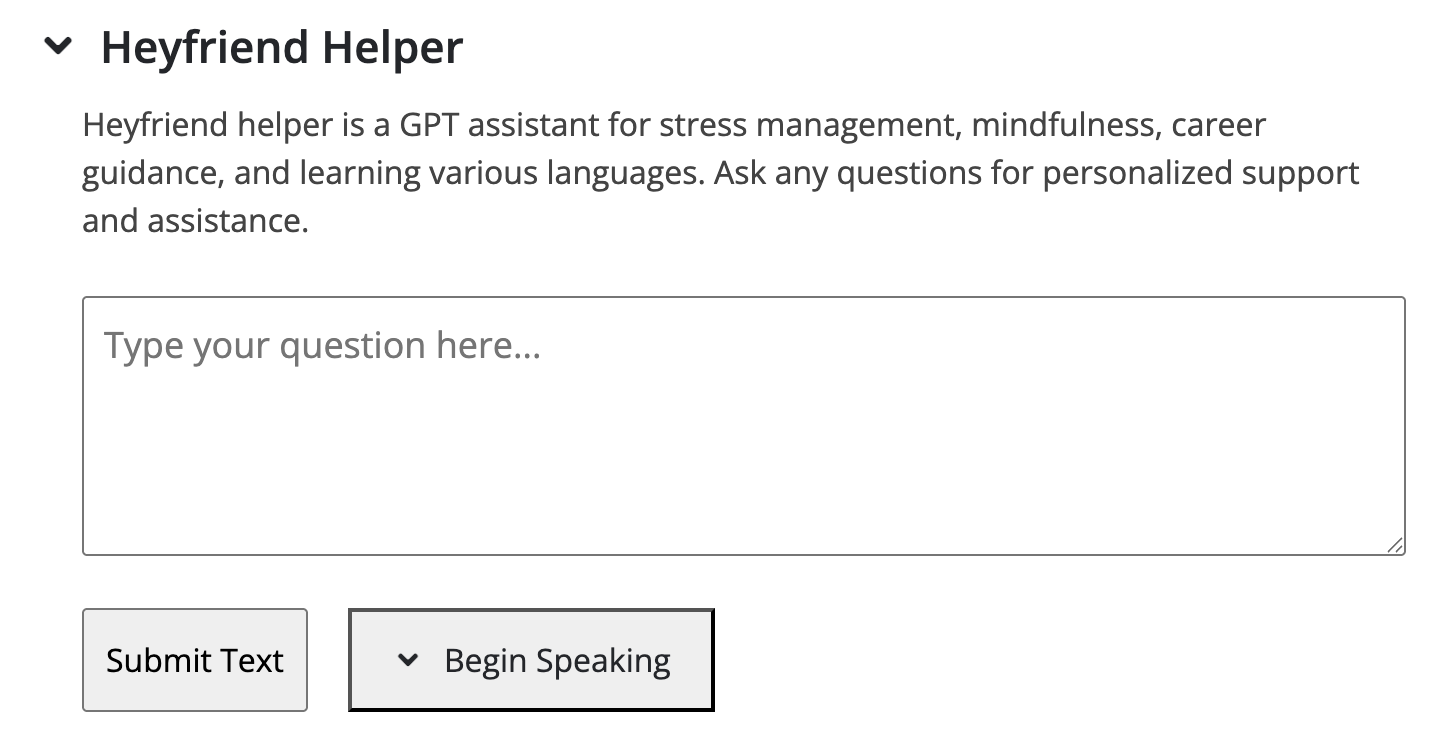}
    \caption{HeyFriend Helper Assistant Interface}
    \label{fig:hf-helper}
\end{figure}

We implemented our conversational AI system using OpenAI’s Assistants API, directly integrating assistant responses within our website rather than relying on a platform-hosted custom-GPT interface. A \textit{HeyFriend Helper} instance can be found within each module of the web-app. An example of the interface is shown in Fig.~\ref{fig:hf-helper}, where users can type or speak their questions to the \system. By embedding the assistant within the platform, users can access this support resource without navigating to external tools or services. This design reduces the complexity associated with locating and using support resources, enabling improved accessible interaction within a single environment. 

\subsection{Common Questions}

\begin{table}[t]
    \centering
    \begin{tabular}{ll}
    \textbf{Topic}  & \textbf{Description} \\ \toprule
      Finding and Getting a Job   & Advice for preparing for job interviews and applications \\
      Relationships & Addresses how to manage relationships and be a better spouse/parent\\
      Well-being &  Advice on how to cope with stress  amid challenges and be mindful\\
      Getting Adjusted to a New Place & Provides recommendations for navigating social resources and building community\\
      Community Resources & Provides local resources that provide need-based support\\
      FitBit & Guidance on using the FitBit, which is part of the larger intervention\\
    \end{tabular}
    \caption{Frequently asked questions addressed in the \emph{Common Questions} module of {\system}.}
    \label{tab:placeholder}
\end{table}

The \emph{Common Questions} module provides answers to common user questions across a broad range of everyday topics in an FAQ-style appearance.
These questions were developed by our community partner, who has extensive experience working with the target population. They are designed to address common challenges users face in daily life and to provide guidance that supports overall well-being.

The questions were developed in consultation with our community agency partner and were also used as references for the \textit{HeyFriend Helper} Assistant.
The development of \textit{HeyFriend Helper} was part of a broader intervention study focused on the use of fitness trackers, such as Fitbit technology; therefore, the inclusion of Fitbit-related questions was intended solely to support participant troubleshooting during the study.
The Fitbit category included practical device-use guidance including how to wear, charge, and sync the device, as well as how to access health metrics within the Fitbit mobile application. 

\begin{figure}[t]
    \centering
    
    \hfill
    \begin{subfigure}[t]{0.48\linewidth}
        \centering
        \includegraphics[width=\linewidth]{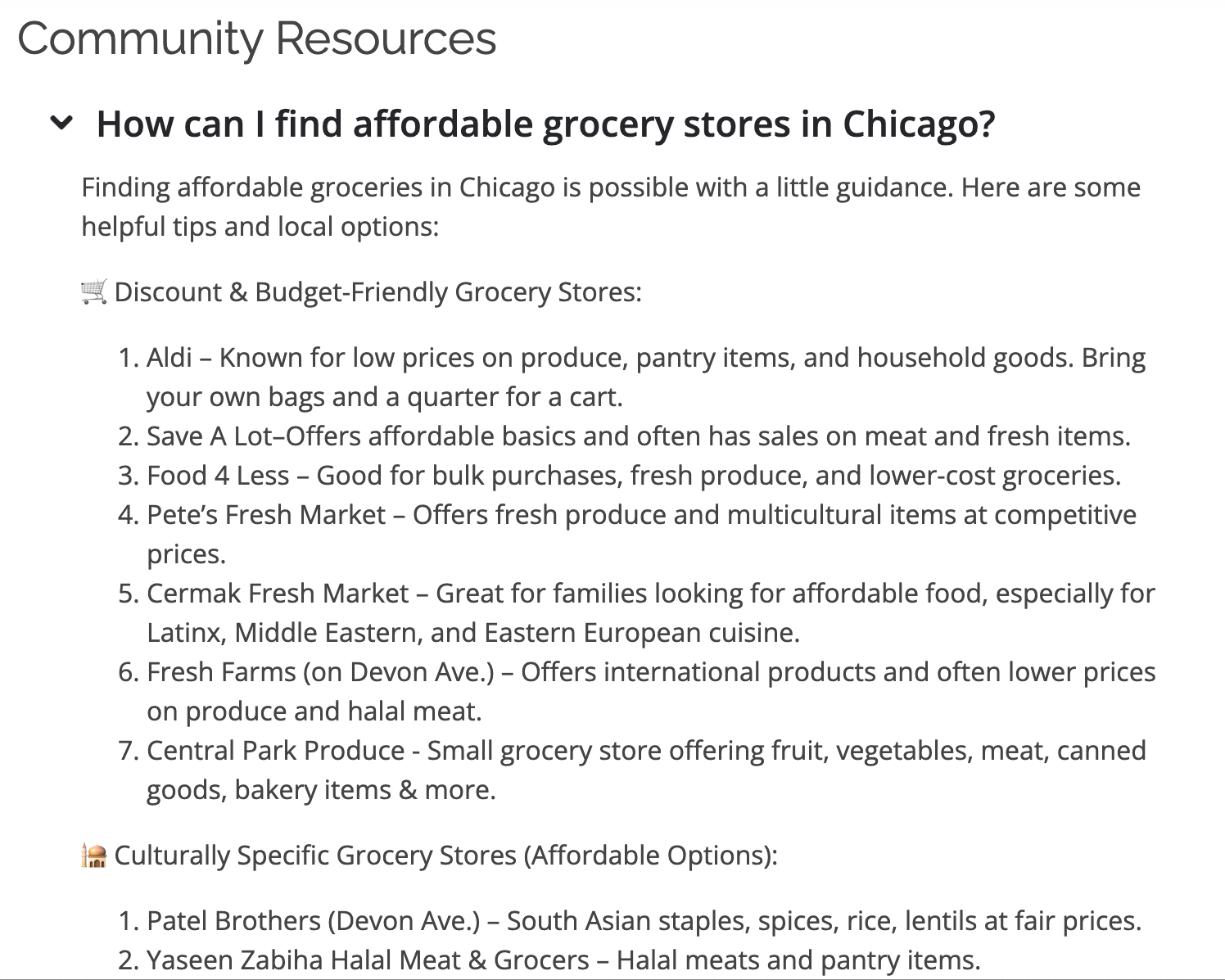}
        \caption{Finding Community Resources}
        \label{fig:cq1-mobile}
    \end{subfigure}
    \begin{subfigure}[t]{0.48\linewidth}
        \centering
        \includegraphics[width=\linewidth]{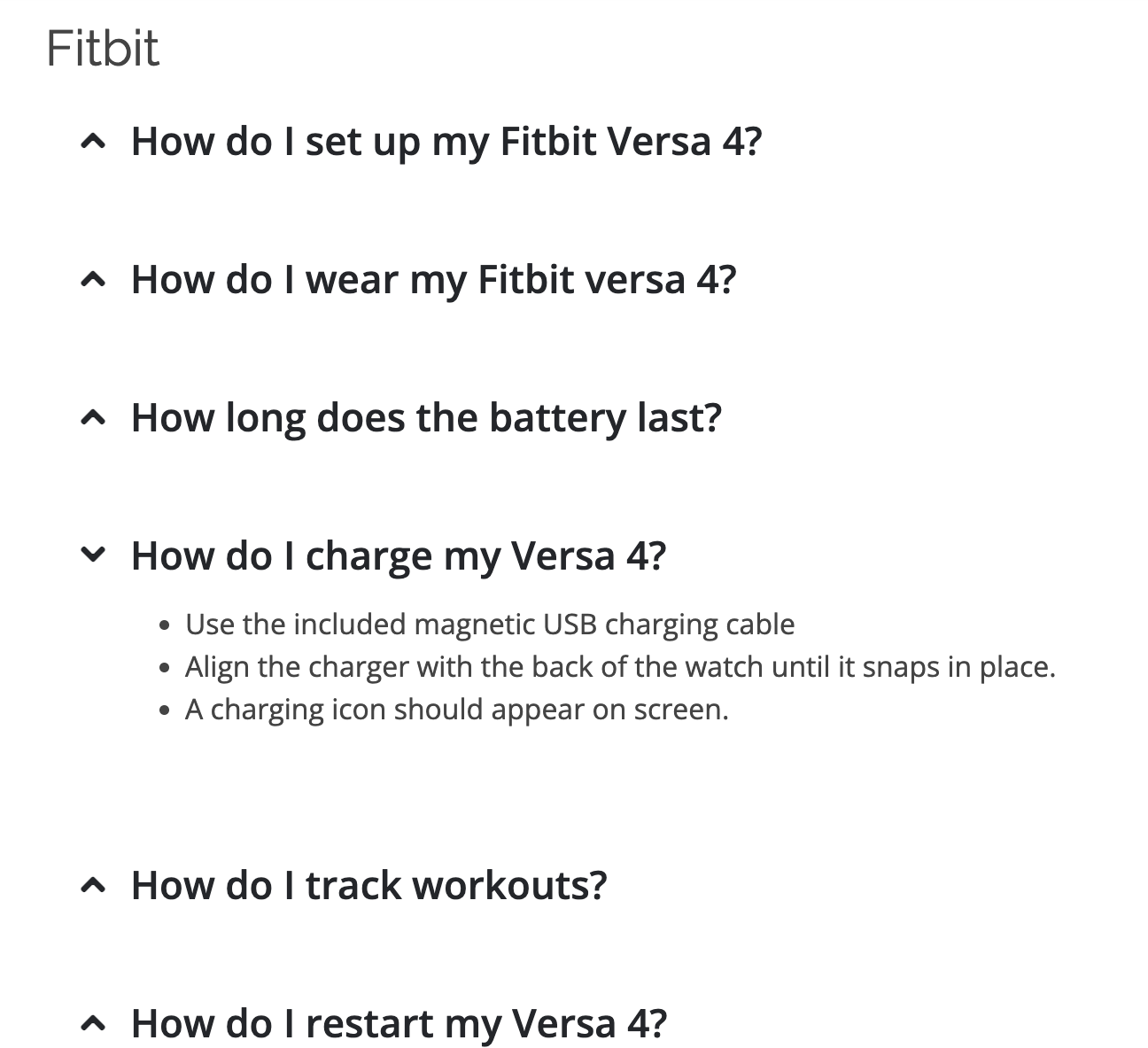}
        \caption{Fitbit Use}
        \label{fig:cq1-expanded}
    \end{subfigure}

    \caption{Common Questions Examples by Category}
    \label{fig:common-quest}
\end{figure}

In addition to general guidance, the section provides practical resources such as curated listings of local Chicago grocery stores categorized by affordability, cultural relevance, and farmers’ markets. It also includes contact information for local agencies offering support services for low-income individuals. Users may browse questions by category and expand individual entries to view their corresponding responses, which can be easily collapsed after review. Seen in Fig. \ref{fig:common-quest}, common questions are displayed by category. Users can view the predefined responses by clicking the question, prompting a drop-down that reveals the answer. 

\begin{figure}[t]
    \centering
    \begin{subfigure}[t]{\linewidth}
        \centering
        \includegraphics[width=0.6\linewidth]{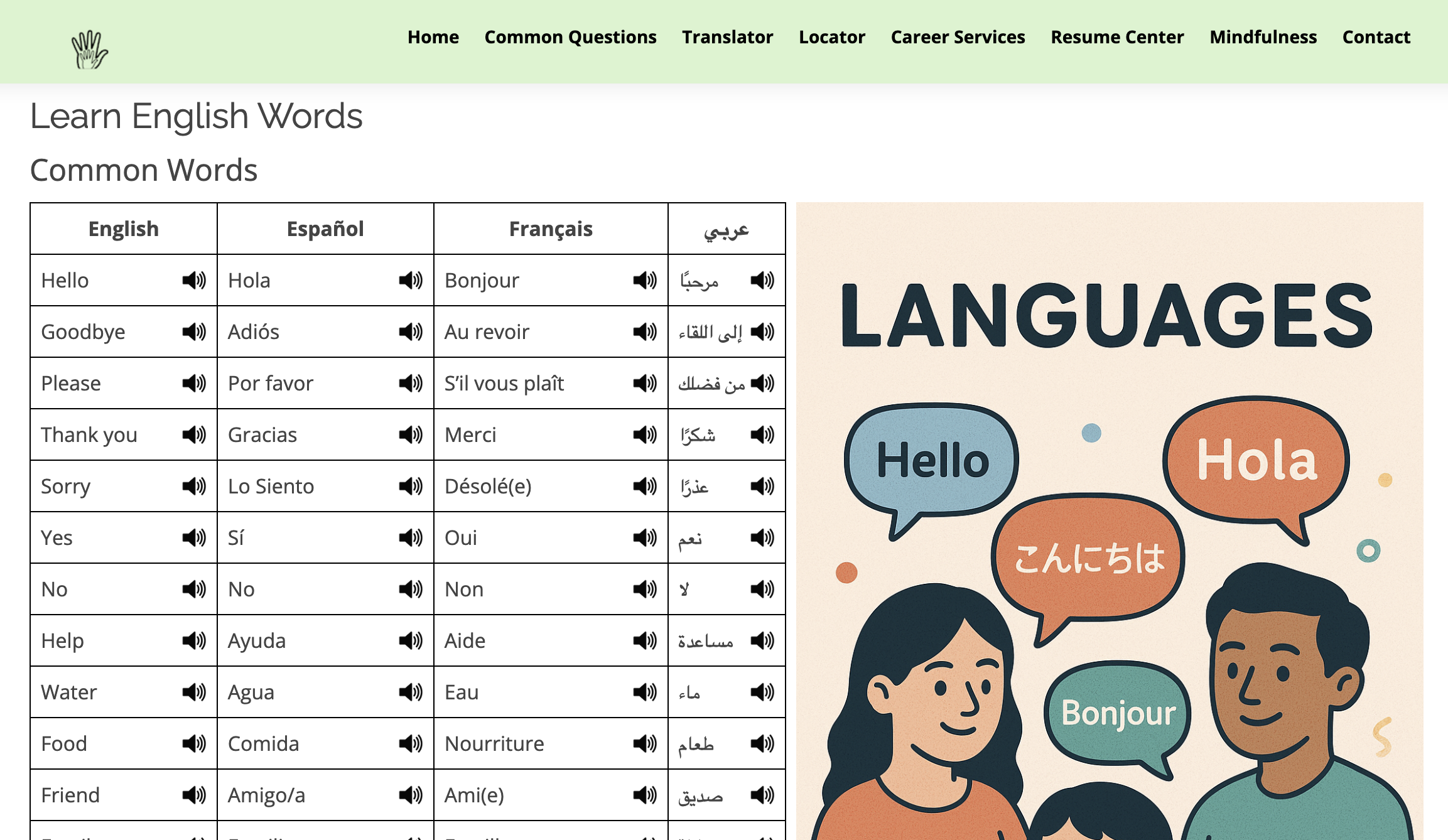}
        \caption{Learn English Feature with Audio Pronunciation and Spelling}
        \label{fig:cq1-desktop}
    \end{subfigure}

    \vspace{0.75em} % optional spacing

    \begin{subfigure}[t]{\linewidth}
        \centering
        \includegraphics[width=0.7\linewidth]{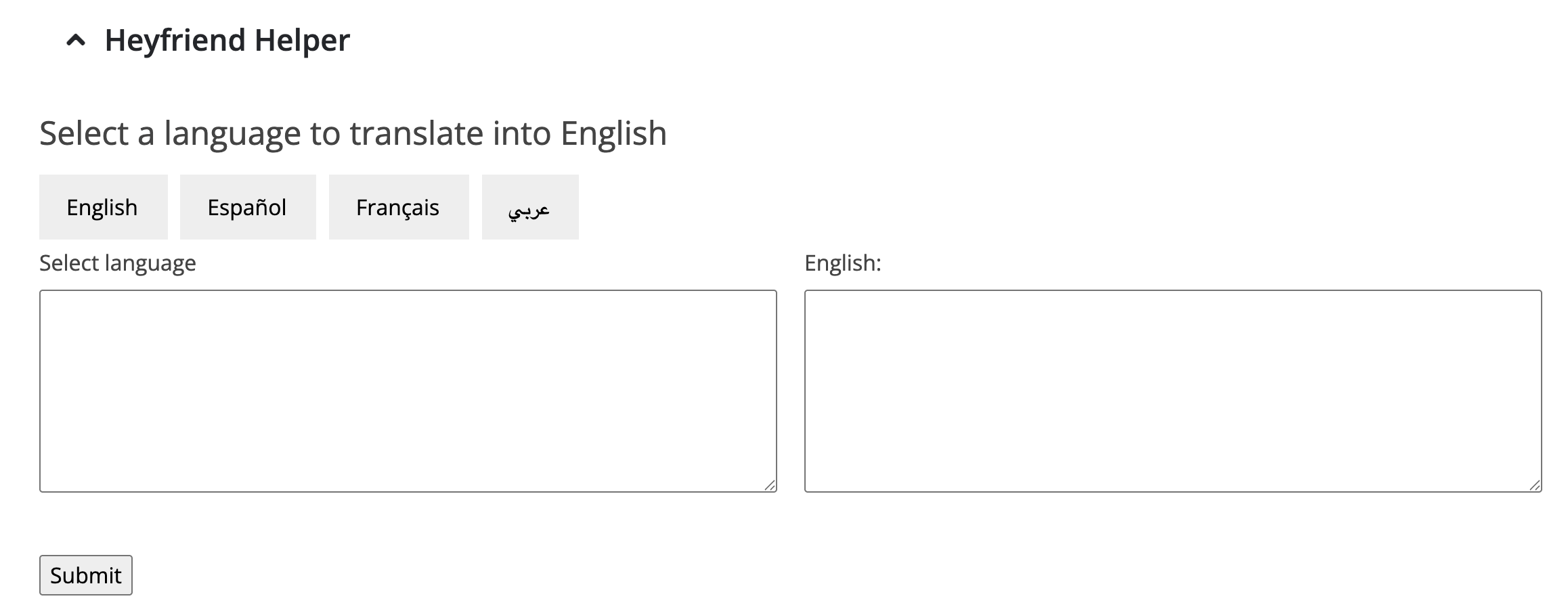}
        \caption{Custom 1:1 Language Translator}
        \label{fig:cq1-mobile}
    \end{subfigure}

    \caption{Common Questions View Examples Across Platforms.}
    \label{fig:translator}
\end{figure}

\subsection{Translator}
This module offers customized language translation and language-learning support. Seen in Fig. \ref{fig:translator}a, we include translations for common words, allowing users to learn new words and pronunciations across languages.  The word categories include: \textit{Common Words},\textit{ Words for Healthy and Unhealthy Relationships}, \textit{Words for Job Search}, \textit{Words for Emotional Well-Being, Words for a Different Kind of Feeling, Greetings, Introductions}, \textit{General Questions and Responses}, \textit{Feeling and Emotional Well-Being}, \textit{Health and Well-Being, and School and Family}. Each sentence or phrase is translated into English, Spanish, French, and Arabic, and each word has its own audio pronoun associated with it. Using Google Translation API \cite{GoogleCloudTranslationAPI}, this module supports 1:1 translation from languages Spanish, French, or Arabic into English, seen in Fig. \ref{fig:translator}b. This feature allows users to enter custom sentences or phrases and receive English translations. These features allow users to learn across multiple languages and practice pronunciation.

\begin{figure}[t]
    \centering

        \includegraphics[width=0.75\linewidth]{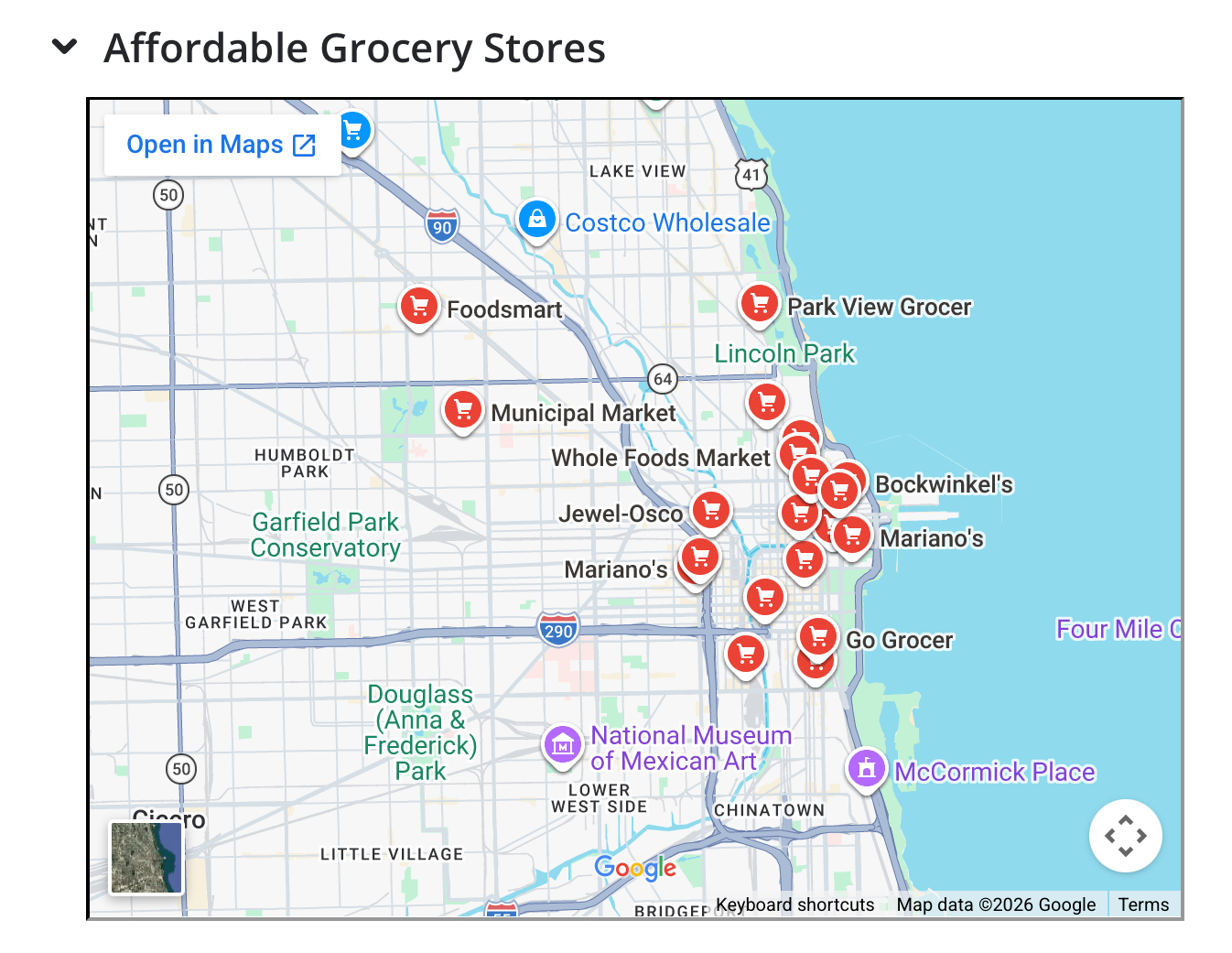}

    \caption{Locator Tool}
    \label{fig:locator-generator}
\end{figure}

\subsection{Locator}
This module provides users with location-based access to nearby resources. Leveraging Google Maps~\cite{GoogleMapsEmbed}, the system displays an interactive map of geographically relevant resources based on the user’s current location. This is done through an explicit embedded search of the form "[resource] near me" (e.g., food pantry near"), using searches to the user's local context. Resource categories include affordable grocery stores, culturally specific grocery stores, farmers’ markets, and food pantries. This approach leverages neighborhood-level relevant searches based on a user's location and result availability. An example of the display can be seen in Fig. \ref{fig:locator-generator}. The results are displayed in an embedded iframe map within the website, which users can expand or collapse to explore nearby stores and resources based on the search.
\begin{figure}[t]
    \centering

    \begin{subfigure}[t]{0.48\linewidth}
        \centering
        \includegraphics[width=\linewidth]{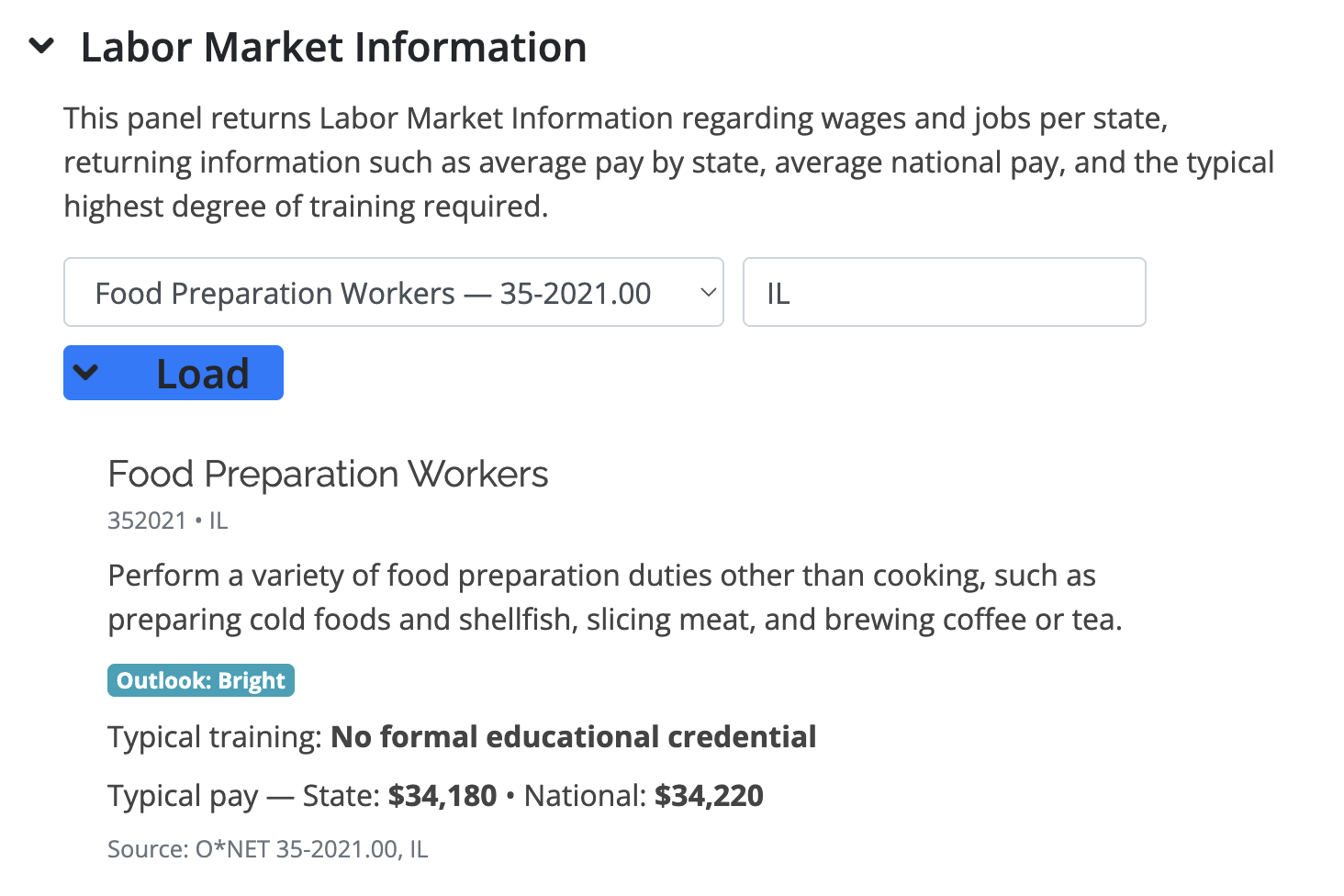}
        \caption{Labor Market Information Module}
        \label{fig:cq1-mobile}
    \end{subfigure}
    \hfill
    \begin{subfigure}[t]{0.48\linewidth}
        \centering
        \includegraphics[width=\linewidth]{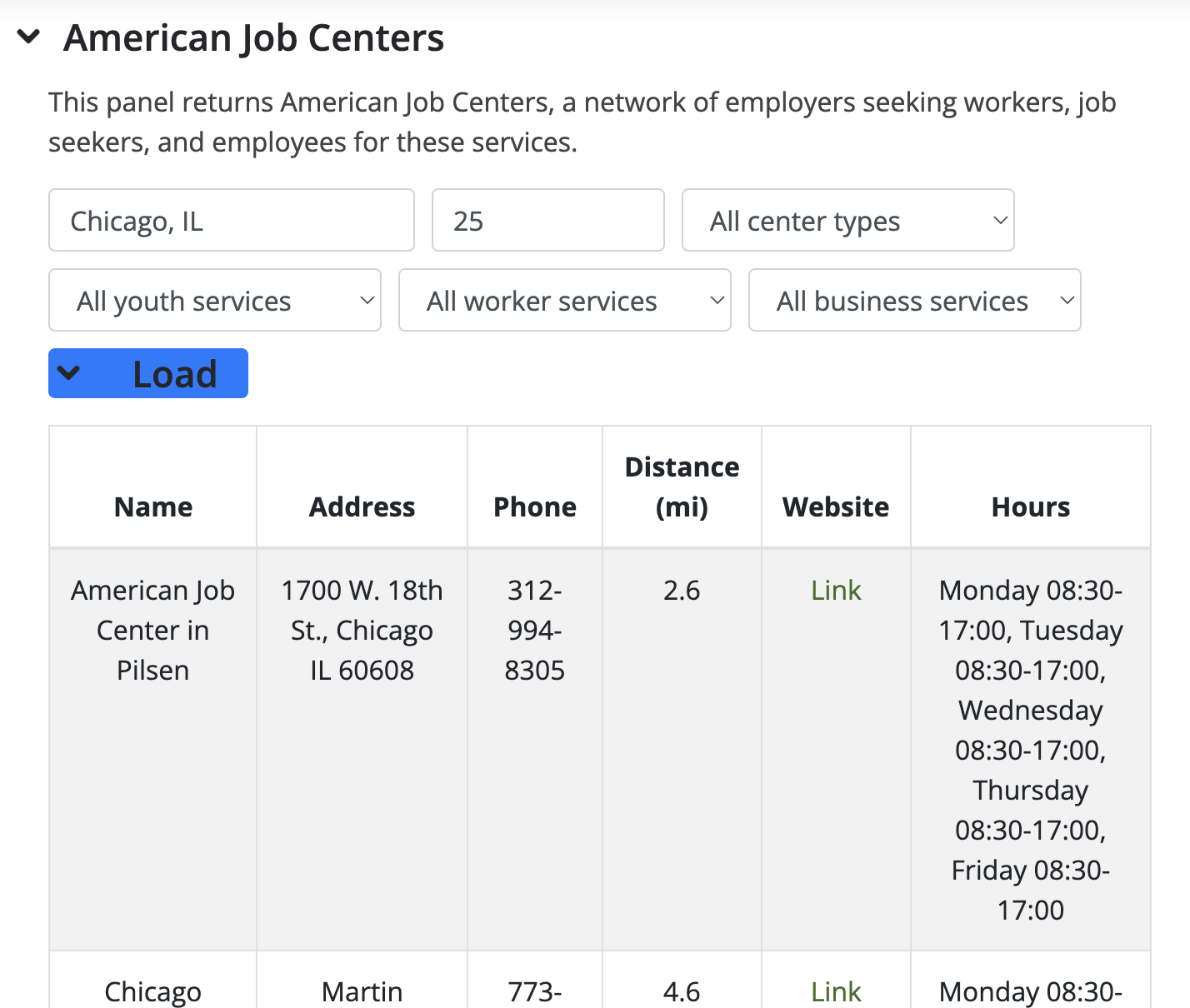}
        \caption{American Job Centers Module}
        \label{fig:cq1-expanded}
    \end{subfigure}

    \caption{Career Services Tool}
    \label{fig:cs-generator}
\end{figure}

\begin{table}[t]
    \centering
    \begin{tabular}{llp{6cm}}
    \textbf{Topic}  & \textbf{Search Parameters} & \textbf{Description} \\ \toprule
      American Job Center   & City, State, or Zip Code & lists locations from US DOL network that provides free career services and training. \\
      Apprenticeship Offices & City, State, or Zip Code  & returns government partners that connect users to apprenticeship programs that combine job training and classroom instruction. \\
      Certifications & O*NET Occupation & lists professional credentials recognized within that occupation that may be required for employment and the accrediting agency. \\
      Employment Patterns & O*NET Occupation & lists the given occupation as a percentage share of workers within industries and est. number of workers employed.\\
      Labor Market Information & O*NET Occupation, State & returns wages and average pay by state and federally.\\
      Occupations & O*NET Occupation, State & returns description of occupation and education \& training required \\
      Occupational Reports & State or U.S. & returns fastest growing occupations, highest paying, most openings, and declining employment by career.\\
      Salaries and Wages & O*NET Occupation, State & returns hourly and annual salary based on location, ranking by percentile.\\
      Skills Gaps & O*NET Occupation & directly compares two occupations to highlight skills, wage, and knowledge differences.\\
      State Resources & State, radius (mi) & provides information on state resources targeted toward unemployed individuals. \\
      Tools and Technology & O*NET Occupation & lists the types of tools and technologies used across vcarious occupations.\\
      Training & City, State, or Zip Code & lists training programs within a distance of location.\\
      Unemployment & State & lists the unemployment rate and count by state.\\
      Youth Programs & City, State, or Zip Code & lists available youth programs and their contact information by local area. \\
    \end{tabular}
    \caption{Description of Each Table in Career Services}
    \label{tab:career-servieces}
\end{table}

\subsection{Career Services}

The \emph{Career Services} module aims to provide support for job search to our study participants.
The resources in this module are targeted toward our study's population of low income communities with comparatively low literacy and education levels.
We present information about job centers, labor markets, and job training in language that is appropriate to the target population.
Our module's presentation focuses on fifty relevant career pathways that are accessible to our target population based on their background.
Our community partners helped us select the fifty most relevant occupations based on their expertise.

%This module presents information related to employment trends, salary ranges, occupation types, and certifications.
The raw data we use in this module comes from the open source CareerOneStopAPI \cite{careeronestop2019} which incorporates up-to-date trends from the U.S. Department of Labor.
\Cref{tab:career-servieces} lists the type of data we gather from CareerOneStopAPI, including the search parameters presented to users.
%Based on the most current data, we display the following relevant tables: (1) American Job Centers, (2) Apprenticeship Offices, (3) Certifications, (4) Employment Patterns, (5) Labor Market Information, (6) Occupations, (7) Occupational Reports, (8) Salaries and Wages, (9) Skills Gaps, (10) State Resources, (11) Tools and Technology, (12) Training, (13) Unemployment, and (14) Youth Programs.
Users are encouraged to explore each panel in the form of a drop-down.
Most panels allow users to refine searches by location and occupation, providing the opportunity to look at national trends compared to local criteria.
The search by location includes city, ZIP code, or state with a relative radius for location-based information (e.g., nearby American Job Centers or Youth Programs).
Furthermore, users can search by occupation to query for specifications (e.g., projected employment, historical salaries, required certifications).
Occupations are defined by Occupational Information Network (O*NET) keywords \cite{onetonline}, standardized descriptions that contain occupation codes for specified occupations to help distinguish occupation types (e.g., Software Developer, Dental Hygienist, Receptionist, and Information Clerk).

To allow users to query for specific jobs, we implemented a 1-to-1 mapping of O*NET occupation code to a displayed occupation, then sent an API request including this code to CareerOneStopAPI to retrieve a response table.
\Cref{fig:cs-generator} displays an example of the types of queried responses for labor market information, which returns wages and jobs per state by occupation.
The Career Services tool also provides information on training programs by location and on the tools and technologies used by occupation, helping users navigate their prospective employment goals and the skills required to gain employment in their desired field.
We display a subset of the  O*NET codes corresponding to the fifty most appropriate occupations selected by our community partner.

\begin{figure}[t]
    \centering

    \begin{subfigure}[t]{0.48\linewidth}
        \centering
        \includegraphics[width=\linewidth]{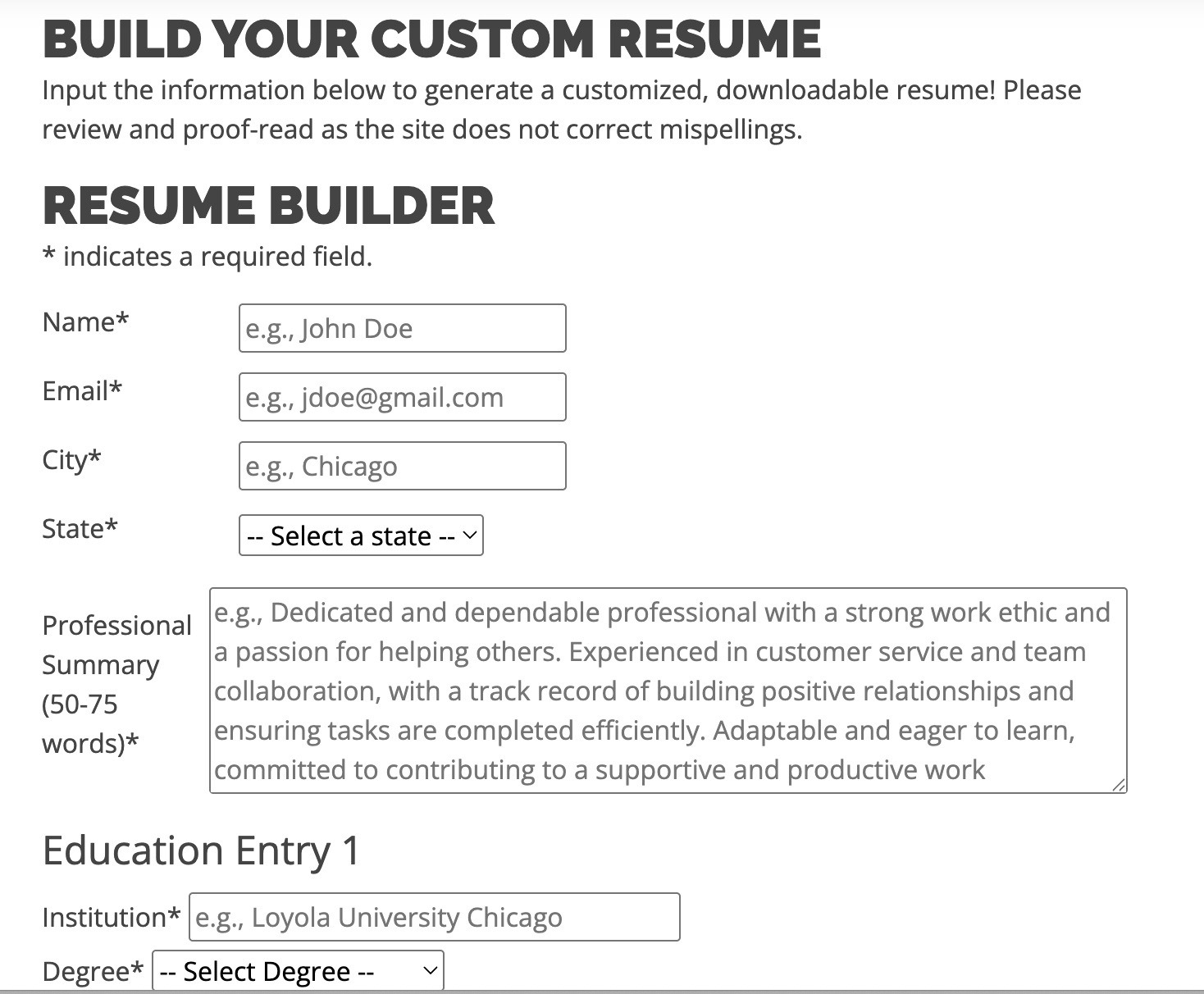}
        \caption{Resume Builder Input}
        \label{fig:cq1-mobile}
    \end{subfigure}
    \hfill
    \begin{subfigure}[t]{0.48\linewidth}
        \centering
        \includegraphics[width=\linewidth]{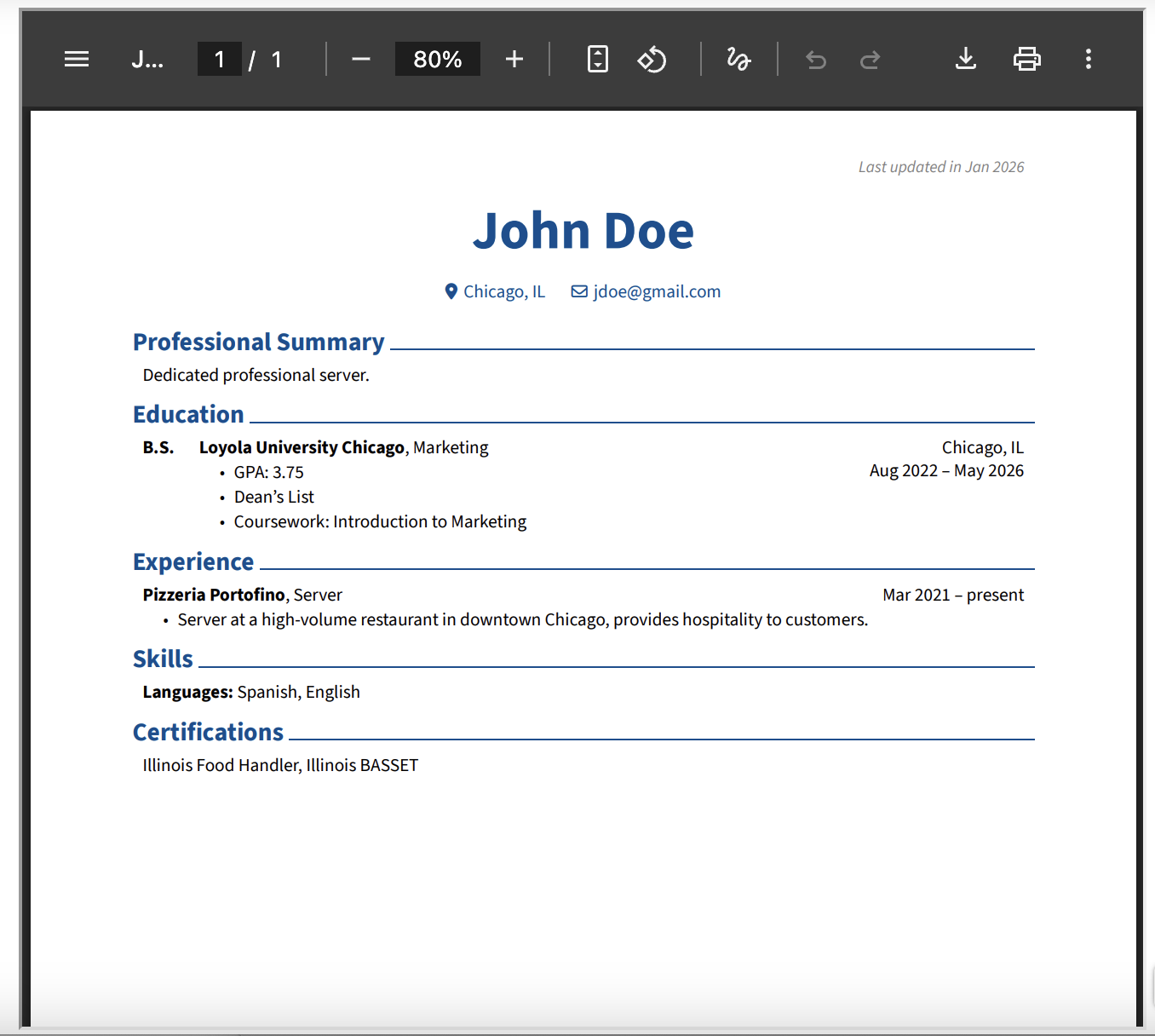}
        \caption{Compiled Resume}
        \label{fig:cq1-expanded}
    \end{subfigure}

    \caption{Resume Builder Implementation}
    \label{fig:resume-generator}
\end{figure}

\subsection{Resume Center}

ResumeGenAI \cite{resumegenai} is an LLM-based resume reviewing tool for job seekers aiming to improve their professional appearances. The system allows users to submit an existing resume for a quantitative scoring and qualitative feedback, compare multiple resume versions, browse templates, and engage in a conversational interface powered by GPT-4o for resume-related guidance. Expanding upon this approach, \textit{HeyFriend Helper} includes a dedicated Resume Center that supports resume creation, written feedback, and interview preparation within a single web-based workflow. The Resume Center also embeds a custom resume builder that enables users to input personal metadata, education, work experience, certifications, and skills through guided form-based inputs to generate a downloadable PDF.

Submitted resume information is sent to a localized helper server as structured JSON, where it is mapped to a formatted YAML representation. Resume layout, formatting, and visual design are handled by \textit{RenderCV}\cite{RenderCV2024}, an open-source, Typst-based resume rendering engine that separates content from presentation through a type-driven schema. \Cref{fig:resume-generator}a showcases the interface design used for collecting user-provided resume information. The helper server invokes RenderCV to compile the YAML representation into a professionally formatted PDF, which is returned to the user as a downloadable document, seen in Fig. \ref{fig:resume-generator}b. This architecture allows \textit{HeyFriend Helper} to provide reproducible, layout-consistent resume generation locally.

Similar to ResumeGenAI, \textit{HeyFriend Helper} includes an LLM-based resume review component that allows users to upload an existing resume and receive structured feedback on its strengths, weaknesses, and areas for improvement.  A key distinction between \textit{HeyFriend Helper} and ResumeGenAI is that our system employs a lightweight, text-driven review pipeline: resume content is extracted directly from uploaded PDFs and passed to GPT-4o to generate narrative, actionable feedback, without performing structured resume scoring, attribute-level evaluation, resume comparison, or template-based analysis. Instead, LLMs in \textit{HeyFriend Helper }are used exclusively for qualitative coaching and guidance rather than quantitative assessment or resume visualization.

We further extend the platform by integrating OpenAI’s speech-to-speech API to support interactive interview practice through voice-based conversations. To make the experience applicable across career types, we selected a set of commonly asked interview questions, including questions such as “\textit{Can you tell me about yourself?}”, “\textit{How do you handle pressure or stressful situations?}”, “\textit{What are your salary expectations?}”, “\textit{What do you do in your leisure time?}”, and “\textit{Do you prefer working independently or on a team?}” Users select a question and then engage in a turn-taking spoken interaction with the system, allowing them to practice articulating responses in a realistic, low-stakes environment. After each speaking turn, the system provides immediate feedback on verbal strengths and areas for improvement, such as clarity, confidence, and completeness of responses. At the end of the session, users receive a written summary of their performance with actionable suggestions to help refine their communication skills. This feature extends beyond text-based support by giving users an opportunity to practice real-time verbal expression, an important tool for interviews and career-readiness. 

\begin{figure}[t]
    \centering

        \includegraphics[width=0.55\linewidth]{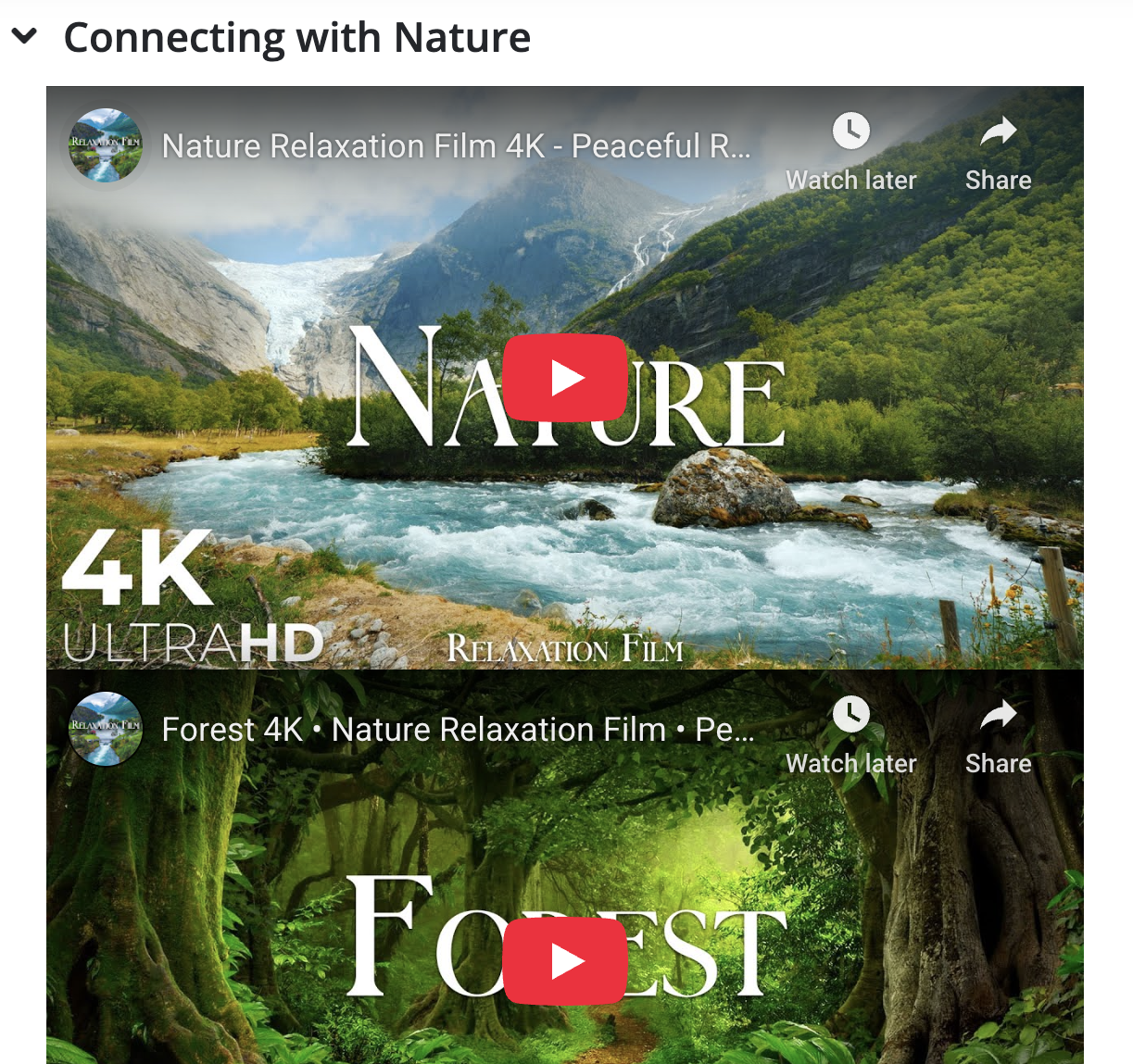}

    \caption{Mindfulness Tool}
    \label{fig:mindfulness}
\end{figure}

\subsection{Mindfulness}

The mindfulness section contains a series of informational publicly available YouTube videos designed to comfort users. Five sub-sections are included in this module: (1) Meditation/Breathing Invitations and Exercises, (2) Wellness, (3) Breathing and Meditation, (4) Connecting with Nature, and (5) Education. An example of the sections and the videos they feature can be seen in Fig. \ref{fig:mindfulness}. Meditation and Breathing Invitations and Exercises features a series of written meditation guides, inviting users to take a moment to ground themselves in breathing techniques to practice meditation. These invitations serve as guides for users and their families to incorporate in their daily lives as they navigate stress and challenges. Our team developed these invitations using previous knowledge and by reviewing best practices around the development of these invitations. 

The remaining sections feature a series of chosen YouTube videos that are embedded into the front-end of the website so users can view them without leaving \textit{HeyFriend Helper}. The Wellness section features relaxing piano sounds and a TedTalk on wellness and social justice. The Breathing and Meditation section features videos on meditative healing sounds and breathing techniques, specifically chosen for users to access calming resources. Next, the Connecting with Nature section includes relaxing nature films aimed at promoting meditation through relaxing visual-audio scenes. Lastly, the Educational section features videos on how to deal with stress and anxiety, how to practice self-compassion, and the science behind meditation techniques. Prior work has shown that short guided meditation videos delivered through platforms such as YouTube can improve mindfulness and occupational well-being \cite{wang2024youtube}. These features, alongside \textit{HeyFriend Helper's} conversational capabilities, serve as a mindfulness hub with resources for users to practice meditation and mindfulness amidst stress and anxiety they may feel.

\section{Results}
The study was conducted with 25 participants in the Midwest over four weeks in a confidential location after receiving Institutional Review Board (IRB) approval. Participant usage was tracked through background JavaScript logging, which captured aggregated interaction data such as buttons clicked, links accessed, and general usage metrics. Qualitative data was collected through semi-structured interviews with 12 participants and a focus group of 7 participants who completed the program. Thematic analysis was with open coding was utilized to analyze the data. 

\subsection{Platform Engagement}

%\todo[inline]{George comment as we seem to have no comment macros: Usage metrics are not the same is whether users find the system useful. I'm really struggling with why we have no information that we can share about whether users found the features useful. Engagement is not a proxy for knowing this. Even if we had some anonymized phrases from participants or a characterization thereof, we would have a much stronger story here. If you don't believe me, consider "number of questions asked". It could be that users asked a lot of questions, only to find the responses not useful.}
% This is resolved: GKT
User activity was measured using a unique session identifier generated for each instance of website use, resulting in 55 distinct user sessions across the study period. Overall, our findings, seen in Table \ref{tab:results-overall}, demonstrate consistent engagement across multiple platform features. Participants actively explored both information and interactive components found within HeyFriend Helper. \Cref{tab:results-tabs} displays the engagement across platform modules. Notably, the Common Questions module was accessed 75 times, suggesting that participants found the predefined questions useful for navigating daily challenges. Furthermore, participants asked 66 questions to the HeyFriend Helper Assistant, including questions such as "\textit{How can I prepare for a dishwashing job?}" and "\textit{How much does a dishwashing job pay?}" in their searches. The distribution of question type can be found in Fig. \ref{fig:question_distribution}. Roughly 30\% (20/66) of the questions to HeyFriend Helper were asked in error or by mistake, indicating that some interactions reflect exploratory use, accidental input, or challenges associated with speech-based and low-literacy interaction contexts. Despite this, the majority of questions were intentional and focused on employment, resume creation, and job preparation, suggesting that participants primarily engaged with the assistant for task-oriented support.

The Resume Building tool was accessed 25 times, with the custom resume generating feature used 17 times, indicating engagement with career-readiness resources and tools. Similarly, the career services feature American Job Center, which lists locations from the US DOL network that offer free career services, was accessed 15 times, indicating that participants used location-based searches for employment tools within the HeyFriend Helper platform. 

Notably, audio pronunciation features, which includes audio pronunciation of different words and phrases across 4 languages, were played 534 times among participants, representing the most frequently used interaction type. This suggests that participants engaged with language learning tools, highlighting the importance of features at improving accessibility.

\begin{table}[t]
    \centering
    \begin{tabular}{ll}
    \textbf{Topic}  & \textbf{Frequency} \\ \toprule
      Number of User Sessions & 55 \\
      Number of questions asked to HF Helper Assistant & 66 \\
      Resumes Generated through Resume Builder & 17 \\
      American Job Center Accessed & 15 \\
      Audio Pronunciation Played & 534 \\

    \end{tabular}
    \caption{Description of Key Results}
    \label{tab:results-overall}
\end{table}

\begin{table}[t]
    \centering
    \begin{tabular}{ll}
    \textbf{Tabs Accessed}  & \textbf{Frequency} \\ \toprule
      Resume & 75 \\
      Career Services & 62 \\
      Mindfulness & 61 \\
      Translator & 54 \\
      Common Questions & 75 \\
      Locator & 45 \\

    \end{tabular}
    \caption{Frequency of Tabs Accessed}
    \label{tab:results-tabs}
\end{table}

\begin{figure}[t]
\centering
\begin{tikzpicture}
\begin{axis}[
    ybar,
    bar width=18pt,
    symbolic x coords={
        Finding a job,
        Resume/CV creation,
        Common-Question type,
        Preparing for an interview,
        Emotional support,
        Questions asked in error
    },
    xtick=data,
    xticklabel style={rotate=35, anchor=east},
    ylabel={Frequency},
    ymin=0,
    nodes near coords,
    height=6cm,
    width=\linewidth
]
\addplot coordinates {
(Finding a job,10)
(Resume/CV creation,15)
(Common-Question type,13)
(Preparing for an interview,5)
(Emotional support,3)
(Questions asked in error,20)
};
\end{axis}
\end{tikzpicture}
\caption{Distribution of question types.}
\label{fig:question_distribution}
\end{figure}
%\todo[inline]{Put as much results in here as we can. Add sections as necessary that correspond to features in \S 3.}
\subsection{Usefulness and Relevance of Job Specificity}

%\todo[inline]{The current analysis (confined to a paragraph) will not fly in the HCI community. There need to be some quotes (anonymized), ideally organized as a table, where we have made some attempt to categorize the comments. This needs to be a lot longer, not shorter.}

The results of the qualitative interviews and focus groups indicated positive feedback on the usefulness of HeyFriend Helper Assistant. Users found HeyFriend Helper useful for job searching, job applications, and interview preparation. Participants used the site to create the objectives portion of their CV and found these features useful. Recognizing the value of CUIs within the HeyFriend Helper website, P1 explained ``\textit{[AI tools] do give [people] easier access to do things, creating resume and finding the answers that they need, but some employers want their resumes a certain way, and is AI gonna give you what that employer wants.}''  
Furthermore, users highlighted the importance of using a variety of tools to search and apply for specific roles. P2 illustrates this noting, ``\textit{Yeah, it's helpful to make a resume and to apply something, or to like to search like jobs or something.  So, this is a very good website,  and I keep searching for jobs, and I'm applying for certain jobs, something with this website. It's very helpful.}'' Users noted that the resume builder tool was helpful for creating a CV and an objective paragraph. 

\subsection{Importance of Career Skills Preparation}
Users appreciated the opportunity to practice interviewing with mock interview questions. P3 expressed, "\textit{[the mock interview questions] kind of gives you more real-time reflection, on what things you can focus, like, where I'm missing, what part do I have to focus more to be closer to my goal.}''  On feature usability and career-readiness, P4 noted, ``\textit{so, just having the tools, especially things like the Resume Builder and mock questions, I don't know, it's just a good way to prepare people, I guess.}''  These responses highlight the value of providing structured opportunities for users to practice and reflect on key career readiness skills within the platform. Participants indicated that having accessible tools for resume development and interview preparation helped them feel more confident and better prepared for real-world applications.

\subsection{Challenges in Instructed Usability Across Features}
Some users noted that HeyFriend Helper's user interface may be difficult for older adults and individuals with limited English proficiency. One challenge of HeyFriend Helper’s integrated, multi-feature design is that its open-ended structure may leave users uncertain about where to begin or how to sequence features effectively (e.g., whether to start with resume development, interview preparation, or conversing with HeyFriend Helper Assistant). Participants recommended creating an introduction video to illustrate Hey Friend Helpers’ functionalities, and easy-to-understand videos on resume writing and other topics that could help non-English speakers. P5 expressed this concern, ``\textit{the resume thing, maybe you can have a checklist where it'll send the participant, like, you know, for example, the first step is, like, research. It'll, like, send a notification out, oh, like, just so you know, like, you should probably do your research soon, and then once you're done with that step, like, oh, start your resume writing.}'' 

\subsection{Need for Accessible Features Across Devices}
 Participants also recommended additional features that could improve the platform’s efficiency, such as personalized, skill-based job searches. While the site currently displays job descriptions based on occupations selected by users, users may be unfamiliar with many of the listed roles. As a result, users often resorted to reviewing job descriptions individually to determine whether each role was a suitable match. Participants described this process as inefficient, particularly given the large number of available occupations, and expressed a preference for a more personalized search approach that could recommend jobs based on a user’s skills, interests, or prior experience. P6 noted, ``\textit{if [HeyFriend Helper] can help, like, on, shortlisting the job description... instead of you going through, oh, this [example], this job is not appropriate for me.}''. 
 
 Furthermore, a common theme was the desire for the HeyFriend Helper's content to be more accessible for individuals who were restricted to only their phones. Some users said that they did not use HeyFriend Helper or used it infrequently because of its format and would have preferred it as an app for easier use, noting that the platform was originally designed for web-based deployment rather than explicit mobile use. Expressing a desire for more versatility, P7 expressed, ``\textit{I feel like if you were to deploy an app, it would definitely be a lot better.'}' Other recommendations from participants, if an app were created, included personalized notifications, such as alerts for relevant job opportunities and reminders of concepts related to classroom instruction on career readiness.

\section{Discussion \& Future Directions}
Future directions for this work include further research into culturally contextualized conversational interfaces that better account for cultural nuance and reduce bias in responses. As LLMs are increasingly used to provide guidance and support, ensuring that interactions are sensitive to diverse social and cultural backgrounds will be essential for improving user experiences and combating bias. This may involve refining instruction design, expanding culturally grounded content, and incorporating community-informed perspectives into system development. Another important extension is the creation of career-specific development features that provide more tailored guidance beyond general employment readiness. For example, future versions could help users identify certifications, training programs, and application-specific steps required for specific career paths. Additionally, strengthening site-specific LLM understanding would allow the assistant to more deeply integrate with existing platform resources, enabling it to reference and guide users to built-in tools (e.g., directing users to the resume builder page during a conversation about job preparation). Together, these directions would move the system toward a more personalized, context-aware, and action-oriented support environment.

\section{Conclusion}

Low-income individuals often face interconnected barriers to employment, including limited access to digital literacy resources, training opportunities, and preparation tools for career readiness. In this work, we introduce \textit{HeyFriend Helper}, a web-based platform that leverages a CUIs to blend multiple forms of support into a single, accessible system tailored to the needs of low-income residents in Chicago. By integrating features such as resume development, interview practice, language learning, well-being resources, and location-based access to community services, the platform demonstrates the potential of conversational AI systems to provide more holistic and personalized employment-readiness support. The results highlight the feasibility of integrating conversational user interfaces with employment and social support resources, and suggest that such systems can improve access to career-readiness tools while supporting broader well-being needs.

\section{Acknowledgments}
We would like to thank Blessing Egbuogu, Samuel Perignon, Maryam Rizvi, Dr. Maria Vidal de Haymes for support with this project. Funding was provided by the Center for Health Outcomes and Informatics Research (CHOIR) Loyola University Chicago.  
\bibliographystyle{ACM-Reference-Format}
\bibliography{reference}


\end{document}